\documentclass[reprint,aps,ams,amsmath,prx,twocolumn,superscriptaddress]{revtex4-1}
\usepackage{graphics}
\usepackage{graphicx}
\usepackage{epsfig}
\usepackage{amsmath}
\usepackage{amssymb}
\usepackage{amsfonts}
\usepackage{bm}
\usepackage{color}
\usepackage{bbm}
\usepackage{hyperref}
 \usepackage[normalem]{ulem}
\usepackage{comment}
\usepackage{soul}

\begin{document}

\title{Hydrodynamic electron transport in graphene Hall-bar devices}

\author{Songci Li}
\affiliation{Department of Physics, University of Wisconsin-Madison, Madison, Wisconsin 53706, USA}

\author{A. V. Andreev}
\affiliation{Department of Physics, University of Washington, Seattle, Washington 98195, USA}
\affiliation{Skolkovo  Institute of  Science  and  Technology,  Moscow,  143026,  Russia}
\affiliation{L. D. Landau Institute for Theoretical Physics, Moscow, 119334 Russia}

\author{Alex Levchenko}
\affiliation{Department of Physics, University of Wisconsin-Madison, Madison, Wisconsin 53706, USA}

\date{April 22 ,2022}

\begin{abstract}
We consider hydrodynamic electron transport in the Hall-bar geometry. The theory is developed for systems with non-Galilean-invariant electron liquids. We show that inhomogeneity of the electron density induced by long-range disorder and gating leads to mixing between the hydrodynamic transport mode and transport relative to the electron liquid. For graphene systems near charge neutrality, these effects lead to strong coupling of the hydrodynamic flow to charge transport. As a result, the effective electrical conductivity of the system may significantly exceed the intrinsic conductivity of the electron liquid. We obtain analytic expressions for the thermoelectric transport coefficients of the system as a function of density in the full crossover region between clean and disorder-dominated regimes.
\end{abstract}

\maketitle

\section{Introduction}

Hydrodynamic transport phenomena of electron liquids in solids continue to attract significant attention; see recent reviews~\cite{Spivak-RMP,NGMS,Lucas-Fong,Polini-Geim,ALJS} and references therein. The required physical conditions for hydrodynamic electron behavior in metals were formulated a long time ago by Gurzhi~\cite{Gurzhi-JETP}: the relaxation length due to momentum-conserving electron-electron (e-e) collisions must be small as compared to the relaxation lengths due to momentum-relaxing electron-impurity (e-i), umklapp (e-e), and electron-phonon (e-p) collisions. Gurzhi also predicted a number of hydrodynamic transport effects in various experimentally relevant geometries~\cite{Gurzhi-UFN}. Experimental realization of the hydrodynamic regime required overcoming serious challenges in sample preparation, and
it took several decades for these effects to be observed in groundbreaking experiments \cite{LWM1,LWM2}.

One of the most striking signatures of hydrodynamic behavior predicted by Gurzhi was the decrease of the resistivity of the systems with increasing temperature, $\partial\rho/\partial T<0$. The reason is that in clean systems, momentum relaxation occurs at the sample boundaries, and the motion of electron fluid takes the form of the Poiseuille flow. Thus the resistivity becomes proportional to the viscosity of the electron liquid, $\rho\propto\eta$. In the Fermi-liquid regime~\cite{Abrikos-Khalat}, the viscosity is proportional to the e-e mean free path, $\eta\propto l_{\text{ee}}\propto 1/T^2$, which immediately yields $\partial\rho/\partial T<0$. The Gurzhi effect, $\partial\rho/\partial T<0$, was  observed in experiments \cite{LWM1,LWM2}.  It is worth noting that the viscous contribution to the resistivity also breaks the Matthiessen's rule (see, e.g., Ref.~\cite{Abrikosov-Book} for detailed discussions), which states that the resistivity $\rho=\sum_i \rho_i$ is proportional to the sum of the partial momentum relaxation rates due to various collision types. 

The recent renewal of interest in hydrodynamic electron transport is largely caused by the significant progress in the fabrication of high-mobility and low-density semiconductor heterostructures, and the advent of boron nitride encapsulated graphene devices has opened new avenues for the exploration of electron hydrodynamic effects~\cite{Spivak-RMP,Polini-Geim}. Novel hydrodynamic manifestations as well as physical regimes of strongly correlated electron liquids \cite{AKS} and Dirac fluids in graphene \cite{Fritz} can be realized in these systems. For instance, the collective character of the viscous motion in finite geometries was predicted to lead to current whirlpools concomitant with the negative nonlocal resistance~\cite{Torre,Levitov-Falkovich,Falkovich-Levitov}. These predictions have found their experimental verification in Refs.~\cite{Bandurin-1,Bandurin-2}. Most recently, SQUID-on-tip imaging technique~\cite{Zeldov} confirmed the existence of these current vortices. It should be mentioned that earlier magnetic sensing and scanning probes provided direct evidence of the Poiseuille flow of electrons in narrow channels~\cite{Ensslin,Sulpizio,Ku,Jenkins,Ilani}. Viscous conductance was also predicted to exceed the ballistic Landauer-Sharvin limit in point contacts~\cite{Guo-1}. This effect was observed in graphene microconstrictions~\cite{Kumar} as well as mesoscale flakes with engineered large-scale defects~\cite{Brar}. The fact that correlations of disorder and e-e interactions may enhance conductivity in the hydrodynamic regime was also predicted to occur in bulk samples~\cite{Hruska,Pal,Guo-2,LLA}. 

Another essential aspect of the problem of electron transport in modern graphene nanodevices is that the electron liquid in them does not possess Galilean invariance. Because of this, in addition to the convective current proportional to the hydrodynamic velocity, such liquids can transport a dissipative current relative to the liquid. Apart from the thermal conductivity $\kappa$, and shear and bulk viscsities, the dissipative properties of such liquids are characterized by the intrinsic conductivity $\sigma$ and the thermoelectric coefficient $\gamma$. The difference between such liquids and Galilean-invariant liquids becomes especially striking at charge neutrality, where charge transport becomes completely decoupled from the hydrodynamic flow \cite{Aleiner,NGTSM,Lucas,Xie-Foster,Principi-2DM,Narozhny-Gornyi}. Experimentally, this manifests in  strong violation of the Wiedemann-Franz (WF) law. The experimentally observed \cite{Crossno}  Lorenz ratio, $L=\kappa\rho/T$, exceeds the universal WF value, $L_\text{WF}=\pi^2/3e^2$, by up to an order of magnitude. Similarly, the magnitude of the observed Seebeck coefficient~\cite{Ghahari} significantly exceeds the prediction of the Mott relation between thermopower and the conductance.

The manifestations of hydrodynamic behavior sensitively depend on the device shape and properties of the electron liquid~\cite{Smet,Hamilton},  which can be accurately tuned in gate-controlled two-dimensional devices. For example, the viscous flow of electrons in the Corbino geometry creates discontinuities in voltage and temperature at the sample boundary \cite{Shavit,LLA-Corbino}, which are caused by the force expulsion from the bulk flow.

In this paper, we consider thermoelectric transport in the Hall-bar geometry in the hydrodynamic regime. Having in mind applications to transport experiments in high-mobility graphene devices, we do not assume Galilean invariance of the electron liquid, and assume the correlation radius of the disorder potential to be long. We study the crossover between the  disorder-dominated regime, where the momentum relaxation occurs in the bulk of the systems, and the clean limit, where the momentum relaxation occurs at the system boundary. We also study the effects of inhomogeneity of gate-modulated electron density on the hydrodynamic flow. 


\section{Hydrodynamic thermoelectric transport}

The hydrodynamic description of fluid dynamics applies to temporal and spatial scales that are large compared to microscopic time and length scales of interparticle collisions \cite{LL-V6}. At such scales, electron liquid locally equilibrates and can be described by the densities of conserved quantities, i.e., particle density $n$, entropy density $s$, and momentum density $\bm{p}$. Having in mind the calculation of the linear response transport characteristics, we need an evolution equation for the momentum density that takes the usual form        
\begin{equation}\label{eq:dp}
\partial_t\bm{p}=-\bm{\nabla}\hat{\Pi}-en\bm{\nabla}\phi,
\end{equation}
where, $\phi$ is the electric potential related to the electron density by the Poisson equation. Here $\hat{\Pi}$ denotes the momentum flux  tensor of the electron liquid,
\begin{equation}
\Pi_{ij}=P\delta_{ij}-\Sigma_{ij}
\end{equation}
that comprises the local hydrodynamic pressure $P$ and viscous stress tensor, 
\begin{equation}
\Sigma_{ij}=\eta(\partial_iu_j+\partial_ju_i)+(\zeta-\eta)\delta_{ij}\partial_ku_k,
\end{equation}
where $\eta$ and $\zeta$ are shear and bulk viscosities, respectively, and $\bm{u}$ is the local hydrodynamic velocity. The fluxes of electrical ($\bm{j}_e$) and entropy ($\bm{j}_s$) currents are given, respectively, 
\begin{subequations}\label{eq:je-js}
\begin{eqnarray}
\bm{j}_e=en\bm{u}+\sigma\bm{\mathcal{E}}-\frac{\gamma}{T}\bm{\nabla}T, \\ 
\bm{j}_s=s\bm{u}-\frac{\kappa}{T}\bm{\nabla}T+\frac{\gamma}{T}e\bm{\mathcal{E}}. 
\end{eqnarray}
\end{subequations}
For the Galilean-invariant systems, the intrinsic conductivity $\sigma$ and thermoelectric coefficient $\gamma$ vanish. In that case, currents are defined only by a macroscopic flow of the system and the thermal conductivity $\kappa$. Note that the electric field $\bm{\mathcal{E}}=-\bm{\nabla}\phi-\frac{1}{e}\bm{\nabla}\mu$ includes the gradient of the local chemical potential. 

In the steady state, $\partial_t\bm{p}=0$, so that with the help of the thermodynamic relation $\bm{\nabla}P=n\bm{\nabla}\mu+s\bm{\nabla}T$~\cite{LL-V5}, 
the equation of motion~\eqref{eq:dp} can be brought to the form of the force balance condition between viscous stresses and driving forces generated by the electric field and temperature gradient. The solution of this Navier-Stokes equation in a given geometry, coupled with proper boundary conditions, defines the spatial profile for the hydrodynamic velocity $\bm{u}(\bm{r})$ to linear order in $\bm{\mathcal{E}}$ and $\bm{\nabla}T$. The knowledge of $\bm{u}(\bm{r})$ defines the currents via Eq.~\eqref{eq:je-js} and proportionality coefficients between them and driving forces yields a matrix of effective thermoelectric coefficients. In the following, we carry out calculations for the rectangular geometry of the Hall bar. 


\section{Clean system with uniform density}
\label{sec:clean_uniform}

Consider a hydrodynamic flow of electron liquid along the $x$ axis in a long strip occupying the region $|y|< d$, so that the total width of the channel is $2d$. To set the stage, let us first consider a clean sample with rough boundaries. In this case, the momentum relaxation occurs via the outflow to the edges of the strip, which is mediated by the viscous stresses.  We assume no-slip boundary conditions for simplicity. The flow velocity is directed along the $x$ axis. Denoting it by $u (y)$ and  projecting the force balance equation to this geometry, we get  
\begin{equation}\label{eq:NS-Hall-bar}
\eta\partial^2_yu(y)=\vec{x}^{\mathbb{T}}\vec{X}. 
\end{equation}
To make the formulas more compact, we introduced column vectors of particle and entropy densities along with thermodynamically conjugated forces \cite{LL-V5},
\begin{equation}\label{eq:x-X}
\vec{x}=\left(\begin{array}{c}n \\ s \end{array}\right), \quad \vec{X}=\left(\begin{array}{c}-e\mathcal{E} \\ \nabla T \end{array}\right).
\end{equation}
The superscript $\mathbb{T}$ denotes transposition in $2\times 2$  column-vector space. The solution for $u(y)$ with the no-slip boundary condition $u(\pm d)=0$ corresponds to the Poiseuille flow,  
\begin{equation}\label{eq:u}
u(y)=\frac{\vec{x}^{\mathbb{T}}\vec{X}}{2\eta}(y^2-d^2).
\end{equation}
For the spatially averaged current densities, we get, from Eq. \eqref{eq:je-js},
\begin{equation}\label{eq:J}
\vec{J}=\frac{1}{2d}\int^{+d}_{-d}\left(\vec{x}u-\hat{\Upsilon}\vec{X}\right)\mathrm{d}y
\end{equation}
where we introduced the column vector of currents $\vec{J}^{\mathbb{T}}=(j,j_s)$, so that $j_e=ej$, and a matrix of the intrinsic  kinetic coefficients of the electron liquid,
\begin{equation}\label{eq:Upsilon}
\hat{\Upsilon}=\left(\begin{array}{cc}\sigma/e^2 & \gamma/T \\ \gamma/T & \kappa/T\end{array}\right). 
\end{equation}
Inserting Eq.~\eqref{eq:u} into~\eqref{eq:J} one gets $\vec{J} = \hat{\Upsilon}_{\mathrm{e}} \vec{X}$, where the effective thermoelectric conductivity matrix of the system is given by
\begin{equation}
    \label{eq:Upsilon_eff_clean}
    \hat{\Upsilon}_{\mathrm{e}} = \hat{\Upsilon} + \frac{d^2}{3\eta} \vec{x}\otimes \vec{x}^{\mathbb{T}}.
\end{equation}
The second term above represents  the contribution of the hydrodynamic transport mode. It is added to the intrinsic thermoelectric conductivity of the liquid (first terms above). The notation $\vec{a}\otimes\vec{b}^{\mathbb{T}}$ is used to denote the direct product of two vectors.  

Setting $\nabla T=0$, one finds a linear relation between the electric current and field $\rho_\text{e}j_e=\mathcal{E}$ that defines an effective electrical resistivity,  
\begin{equation}\label{eq:rho-clean}
\rho_{\text{e}}=\frac{\sigma^{-1}}{1+(e^2/3\sigma\eta)(nd)^2}.
\end{equation}
We see that the resistivity has a Lorentzian shape as a function of density. The temperature dependence of its width is implicit and defined by the product of intrinsic conductivity and viscosity. For graphene in the regime close to charge neutrality, $\sigma(T)$ is weakly (logarithmically) temperature dependent and is of the order of quantum conductance \cite{Mishchenko,Kashuba,FSMS}. In the same regime, also modulo logarithmic renormalizations, viscosity behaves as $\eta(T)\sim (T/v)^2$ \cite{Fritz}, where $v$ is the band velocity in graphene. Therefore, to the main approximation, the width of the Lorentzian grows linearly with the increase of temperature.      

Repeating the same steps, but setting the total electrical current to zero, $j_e=0$, gives a unique relation between the electric field and temperature gradient. This determines an effective thermopower (Seebeck coefficient) in the form  
\begin{equation}\label{eq:Q-clean}
Q_{\text{e}}=\frac{e\mathcal{E}}{\nabla T}=\frac{\gamma/T+nsd^2/3\eta}{\sigma/e^2+(nd)^2/3\eta}.
\end{equation}
Finally, using this result in the second column of Eq.~\eqref{eq:J}, one finds an effective thermal conductivity,
\begin{equation}\label{eq:kappa-clean}
\kappa_{\text{e}}=\kappa+T\frac{(sd)^2}{3\eta}-T\frac{(\gamma/T+nsd^2/3\eta)^2}{\sigma/e^2+(nd)^2/3\eta}.
\end{equation}
We remind the reader that the thermal conductivity is defined as a proportionality coefficient between the entropy current and temperature gradient at vanishing electrical current, namely, $\kappa=-(Tj_s/\nabla T)_{j_e=0}$. 

From Eqs.~\eqref{eq:rho-clean}--\eqref{eq:kappa-clean}, it becomes clear that viscous effects significantly modify the thermoelectric properties of the system. In a generic strongly correlated system, the density and temperature dependence of the pristine coefficients in the thermoelectric matrix $\hat{\Upsilon}(n,T)$ is rather complex. However, close to charge neutrality, one may simplify the main results and bring them to a rather universal form. Indeed, in the regime of low doping $n/s\ll1$, the off-diagonal elements in Eq. \eqref{eq:Upsilon} may be estimated to scale as $\gamma/T\sim n/s\ll1$. Therefore, if the strip is sufficiently wide, such that $s^2 d^2/\eta \gg 1$,
the effective thermal conductivity from Eq.~\eqref{eq:kappa-clean} simplifies to 
\begin{equation}\label{eq:kappa-clean-simple}
\kappa_\text{e}\approx \kappa +  \frac{\sigma}{e^2}
\frac{Ts^2}{n^2+\Gamma^2},\quad \Gamma^2=\frac{\sigma}{e^2}\frac{3\eta}{d^2}.
\end{equation} 
It is of interest to observe that in this low-doping limit, there exists a temperature regime where $\kappa/T\ll (sd)^2/\eta$, and thus the effective thermal conductivity turns out to be practically independent of its pristine value $\kappa$. In the same limit, the thermopower in Eq.~\eqref{eq:Q-clean} takes the form 
\begin{equation}
Q_\text{e}\approx \frac{ns}{n^2+\Gamma^2}. 
\end{equation} 
Lastly, using the resistivity from Eq.~\eqref{eq:rho-clean} and assuming $n\ll s$, we obtain the Lorenz ratio in the form
\begin{equation}\label{eq:L}
L_\text{e}=\frac{\kappa_\text{e}\rho_\text{e}}{T}\approx\frac{1}{e^2}\left(\frac{\Gamma s}{n^2+\Gamma^2}\right)^2. 
\end{equation} 
This formula captures the main characteristics of the observed anomalously large $L(n,T)$. Indeed, the height of the peak in $L_{\mathrm{e}}$ at $n=0$  is set by the ratio of $(s/\Gamma)^2$. Since both viscosity and entropy density scales as $\eta\sim s\sim (T/v)^2$ at low doping, then $L_\mathrm{max}/L_{\text{WF}}\sim (T/E_d)^2$ with the energy scale $E_d=v/d$.
Taking a micron-size strip $d\sim 1\mu$m and $v\sim10^6$m/s, one gets $E_d\sim10$K. As a consequence, for temperatures $T>50$K where the hydrodynamic regime is expected to be established by ensuring $l_\text{ee}\ll d$, one may easily achieve $L_{\text{max}}/L_{\text{WF}}\gg1$. 

\begin{figure*}[t]
\includegraphics[width=0.325\linewidth]{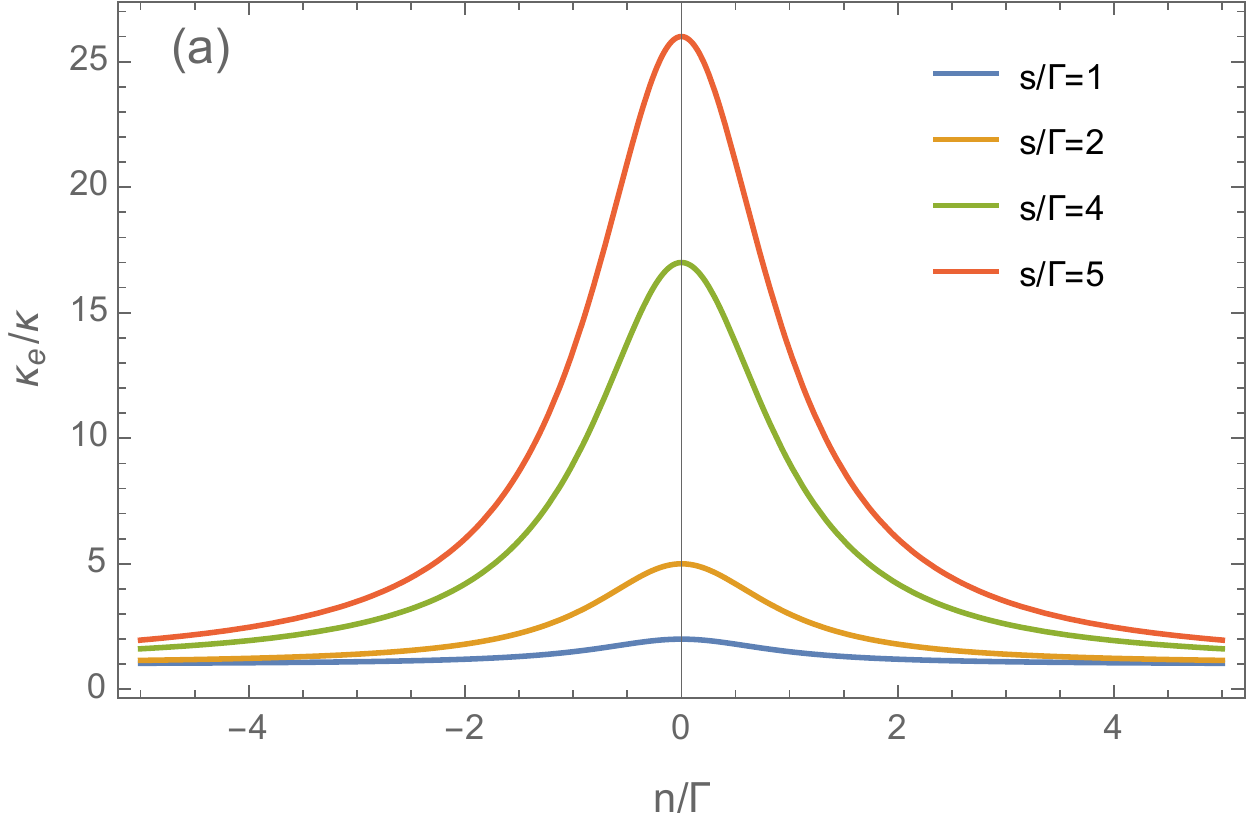}
\includegraphics[width=0.325\linewidth]{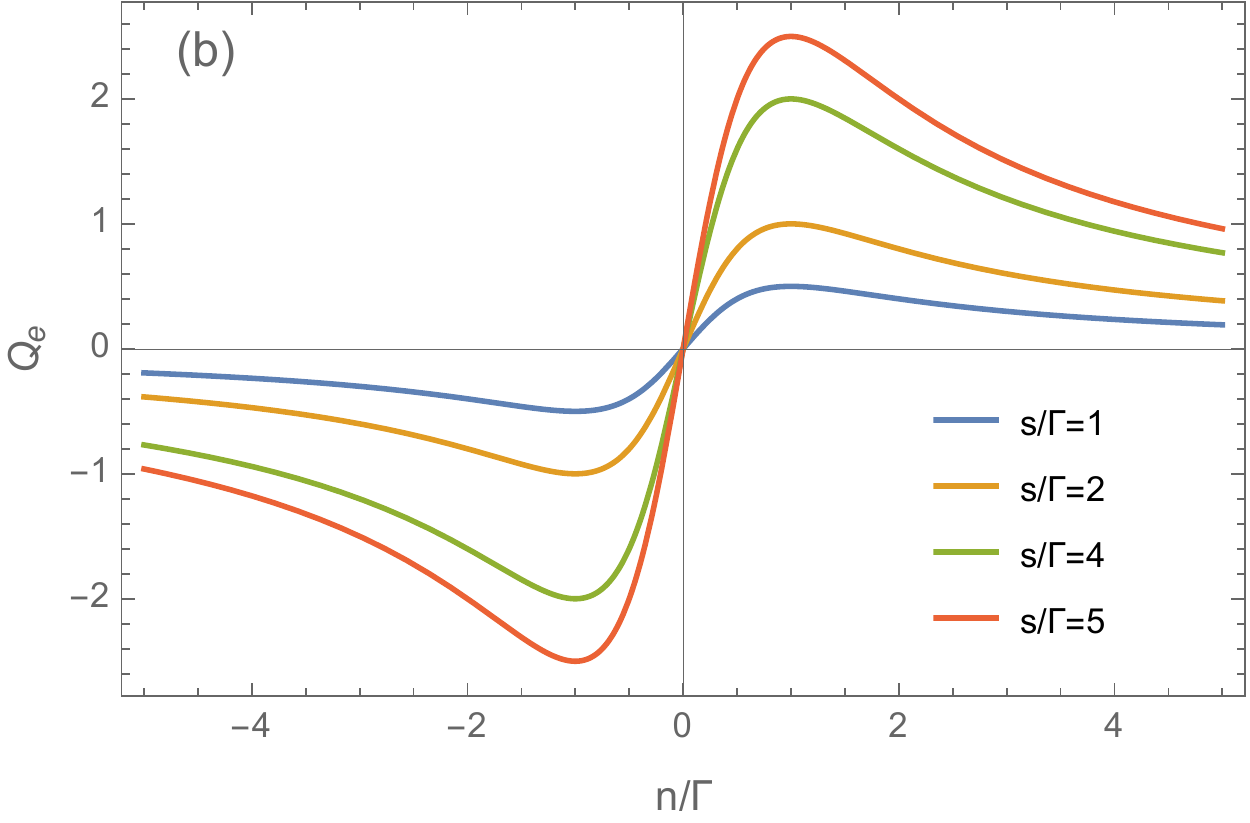}
\includegraphics[width=0.325\linewidth]{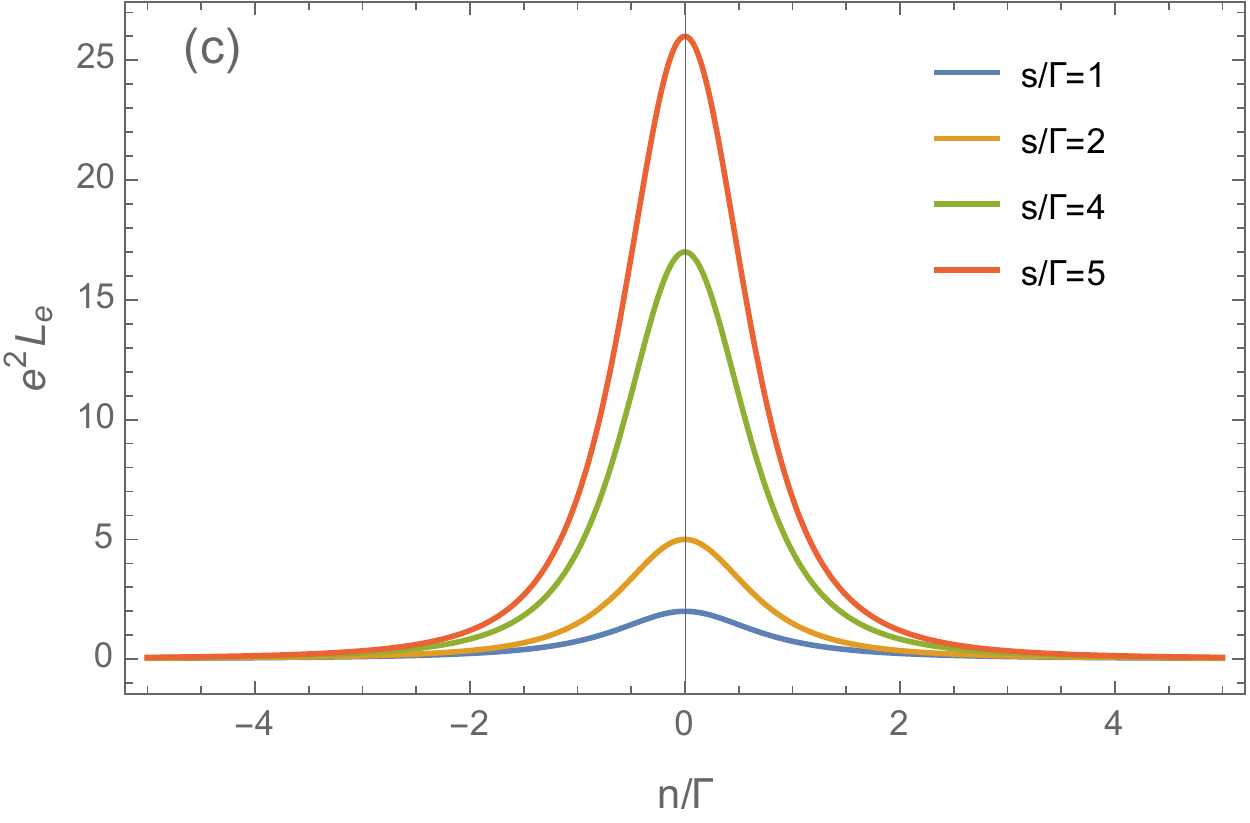}
\caption{(a) Density dependence of the normalized thermal conductivity, (b) thermopower, and (c) Lorenz ratio for several different values of temperature, which is implicit in the entropy to viscosity ratio as captured by the $s/\Gamma$ ratio. In the case of thermal conductivity, the results are also dependent on the additional parameter $T\sigma/\kappa e^2$ that we set to unity for simplicity as it only influences the peak height.}
\label{fig:K-Q-L}
\end{figure*}

The principal findings of this section are summarized in Fig. \ref{fig:K-Q-L}, where we plot $\kappa_{\text{e}}(n,T)$, $Q_\text{e}(n,T)$, and $L_\text{e}(n,T)$ in scaled units of $n/\Gamma$ for different values of $s/\Gamma\propto T/E_d$.

\section{Clean system with nonuniform density}
\label{sec:clean_nonuniform}

The consideration of hydrodynamic transport in clean systems in Sec. \ref{sec:clean_uniform} assumed the electron density to be uniform across the system. Doping by an external gate generally creates an inhomogeneous particle density $n(\bm{r})$ and entropy density $s(\bm{r})$ in the sample. In this section, we consider the effects of nonuniform gating on hydrodynamic thermoelectric transport in clean Hall-bar systems.

In our model geometry, in which the clean Hall is separated from the gate by a uniform distance, the  densities of electrons and entropy will depend on the $y$ coordinate. For a flow in the Hall bar along the $x$ direction, the electromotive force and the temperature gradient, being purely potential vector fields, are spatially uniform. In this case, we get, from Eq.~\eqref{eq:NS-Hall-bar}, 
\begin{equation}
u'(y)=-\int_0^y \mathrm{d}z \frac{1}{\eta(z)} \left[ n(z) e\mathcal{E} - s(z) \nabla T \right].
\end{equation}
Next, using the no-slip boundary condition $u(\pm d)=0$ and introducing the notations
\begin{equation}
\label{eq:N_S_integrated}
\mathcal{N}(y)=\!\int_0^y \mathrm{d}z \int_0^z \mathrm{d}z' \frac{n(z')}{\eta (z')} ,\quad \mathcal{S}(y)=\!\int_0^y \mathrm{d}z \int_0^z \mathrm{d}z' \frac{s(z')}{\eta (z')},
\end{equation}
we find
\begin{equation}
\label{eq:u_Poiseuille_solution}
u(y) = -\left[ \mathcal{N}(y)- \mathcal{N}(d)\right]e\mathcal{E} +  \left[ \mathcal{S}(y)- \mathcal{S}(d)\right] \nabla T.
\end{equation}
The net particle and entropy currents across the sample are thus given by
\begin{align}\label{eq:j_integrated_Poiseuille}
&2d\vec{J}  =  - 2  e\mathcal{E} \int_0^d \vec{x} (y)\left[ \mathcal{N}(y)- \mathcal{N}(d)\right]\mathrm{d}y  \nonumber \\
 &+2 \nabla T \int_0^d\vec{x} (y)  \left[ \mathcal{S}(y)- \mathcal{S}(d)\right]\mathrm{d}y   - 2 \int_0^d \mathrm{d}y \hat{\Upsilon} (y)  \vec{X} .
\end{align}
Integrating by parts in the first two terms, we can express the electric current and heat flux in the form
\begin{align}
\label{eq:electric_current_general}
eI =& 2  \mathcal{E} \left[ e^2  \int_0^d N (y)\mathcal{N}'(y)\mathrm{d}y + \int_0^d  \sigma (y) \mathrm{d}y\right]  \nonumber \\
&- 2 e \nabla T  \left[
\int_0^d N (y)  \mathcal{S}'(y) \mathrm{d}y + \frac{1}{T}
\int_0^d \gamma(y)\mathrm{d}y   \right] , \\
\label{eq:heat_flux_general}
TI_s =& 2  e \mathcal{E} \left[  T  \int_0^d  S (y)\mathcal{N}'(y) \mathrm{d}y+ \int_0^d \gamma(y)\mathrm{d}y \right] \nonumber \\
&- 2  \nabla T  \left[ T
\int_0^d  S (y)\mathcal{S}'(y)\mathrm{d}y +
\int_0^d \kappa (y)   \mathrm{d}y  \right].
\end{align}
Here the quantities  $N(y)$ and $S(y)$  denote the integrated electron density and entropy density, respectively,
\begin{equation}
\label{eq:N_S_def}
N(y) = \int_0^y n(z) \mathrm{d}z, \quad S(y) = \int_0^y s(z) \mathrm{d}z.
\end{equation}
Setting the temperature gradient to zero, we obtain, for the effective conductivity,
\begin{eqnarray}
\label{eq:conductivity_Poiseuille_nonuniform}
\sigma_\text{e}
= \frac{1}{d}\left(   e^2 \int_0^d N (y) \mathcal{N}' (y)\mathrm{d}y+  \int_0^d \sigma(y)\mathrm{d}y \right).
\end{eqnarray}
Due to the linear dependence of $n(y)$ on the gate voltage, Eq.~\eqref{eq:conductivity_Poiseuille_nonuniform} corresponds to a Lorentzian gate voltage dependence of the resistivity. Assuming, for simplicity, that the viscosity is position independent, we get
\begin{eqnarray}
\label{eq:conductivity_Poiseuille_simplified}
\sigma_\text{e} = \frac{1}{d }\left(   \frac{e^2}{\eta} \int_0^d N^2 (y)\mathrm{d}y +  \int_0^d \sigma(y) \mathrm{d}y\right).
\end{eqnarray}
At zero electric current, we obtain the thermopower
\begin{eqnarray}
\label{eq:thermopower_Poiseuille}
Q_\text{e} &=& \frac{e \mathcal{E}}{\nabla T} =
\frac{\int_0^d  N (y)\mathcal{S}'(y)\mathrm{d}y +
\int_0^d\mathrm{d}y\,\gamma (y)/T}{\int_0^d N(y)\mathcal{N}'(y)\mathrm{d}y+  \int_0^d\mathrm{d}y\, \sigma(y)/e^2},
\end{eqnarray}
and the thermal conductivity 
\begin{align}
\label{eq:kappa_eff_Poiseuille}
& \kappa_\text{e} = \frac{1}{d} \left[ T
\int_0^d S (y)  \mathcal{S}'(y)\mathrm{d}y  +
\int_0^d \kappa (y)   \mathrm{d}y  \right. \nonumber \\
&-\left( T
\int_0^d S (y)  \mathcal{N}'(y) \mathrm{d}y +
\int_0^d \gamma (y) \mathrm{d}y \right)  \nonumber \\
&\times
\left.\frac{
\int_0^d N (y)  \mathcal{S}'(y)\mathrm{d}y  +
\int_0^d \mathrm{d}y\, \gamma (y)/T  }{   \int_0^d N (y) \mathcal{N}' (y)\mathrm{d}y+  \int_0^d \mathrm{d}y\, \sigma (y)/e^2 }  \right].
\end{align}
Note that the intrinsic terms scale linearly with the channel width, while the convective ones scale quadratically with $d$. Therefore, in sufficiently wide channels, we may neglect the intrinsic terms. We then get, for the thermal conductivity,
\begin{align}
\label{eq:kappa_eff_wide_channel}
&\kappa_\text{e} = \frac{T}{d} \left[
\int_0^d S (y)  \mathcal{S}'(y) \mathrm{d}y\right. \nonumber \\ 
& \left.- \frac{
\int_0^d \mathrm{d}y N (y)  \mathcal{S}'(y)
\int_0^d \mathrm{d}y S (y)  \mathcal{N}'(y)}{ \int_0^d \mathrm{d}y \,  N (y) \mathcal{N}' (y)+  \int_0^d \mathrm{d}y\, \sigma (y)/e^2}  \right].
\end{align}
The above expressions describe the transport coefficients in clean Hall-bar samples in the presence of inhomogeneous gate doping.

In this model, strong thermoelectric coupling is also expected, along with the large conductivity even when the system is, on average, charge neutral. 
For $n\ll s$ treating electron density $n(y)$ as the sole inhomogeneous quantity, we would get simplified expressions 
\begin{subequations}
\label{eq:transport_clean_inhomogeneous}
\begin{align}\label{eq:sigma_clean_inhomogeneous}
    & \sigma_\text{e} = \sigma + \frac{e^2}{\eta}\int^d_0 \frac{\mathrm{d}y}{d} N^2(y), \\
    \label{eq:Q_clean_inhomogeneous}
    & Q_\text{e} = \frac{\eta\gamma/T + s\int^d_0 \frac{\mathrm{d}y}{d} yN(y)}{\eta\sigma/e^2 + \int^d_0 \frac{\mathrm{d}y}{d} N^2(y)}, \\
    &\kappa_\text{e} =\kappa+ \frac{Ts^2}{\eta}\Bigg(\frac{d^2}{3} - \frac{\left[\int^d_0 \frac{\mathrm{d}y}{d} yN(y)\right]^2}{\frac{\eta\sigma}{e^2}+\int^d_0 \frac{\mathrm{d}y}{d} N^2(y)}\Bigg).\label{eq:kappa_clean_inhomogeneous}
\end{align}
\end{subequations}
These results show that the transport properties of graphene Hall-bar samples depend on the spatial distribution of the electron density over the sample,  not just the spatial average of the electron density. In particular, Eq.~(\ref{eq:sigma_clean_inhomogeneous}) shows that the effective conductivity of a sample that is, on average, charge neutral can significantly exceed the intrinsic conductivity of the electron liquid. This occurs because of local deviations from charge neutrality. The applied external electric field $E_x$ induces a local force  $n(y) e E_x$ on the electron liquid, which generates a vortical hydrodynamic flow in the $x$ direction. The  velocity $u_x (y)$ of this flow is determined by the balance between the viscous force and the driving force, and therefore is correlated with the electron density. As a result, the induced hydrodynamic flow contributes to the effective conductivity of the sample. This contribution is described by the second term in Eq.~(\ref{eq:sigma_clean_inhomogeneous}). The analogous phenomenon for enhancement of the effective bulk conductivity in samples with long-range disorder was discussed in Ref.~\cite{LLA}. We also note that the Peltier coefficient in Eq.~\eqref{eq:Q_clean_inhomogeneous} also depends on the entire distribution of the electron density, and need not turn to zero when the average charge density in the sample vanishes. Therefore, when the macroscopic electron density of the sample is tuned by the gate voltage, the sign reversal of the thermoelectric coefficients does not generally coincide with global charge neutrality. 

The results of this section can be tested by space-resolving thermal imaging probes \cite{Zeldov-Nature,Zeldov-Science} in the search for the hydrodynamic electron transport in graphene, and in the measurements based on the nonlocal noise thermometry \cite{Waissman}. They are also relevant for experimental interpretation of macroscopic transport measurements in graphene Hall bars. 


\section{Effect of long-range disorder}
\label{sec:disorder}

In the clean systems considered in Secs. \ref{sec:clean_uniform} and \ref{sec:clean_nonuniform}, momentum relaxation occurs only due to diffusive scattering at the system boundaries. As a result, the hydrodynamic contribution to the effective thermoelectric conductivity matrix increases with the width of the Hall bar $2d$. In the presence of disorder, momentum relaxation also occurs in the bulk. As the disorder strength increases, the effective conductivity becomes independent of $d$. The crossover between these two regimes was originally discussed by Gurzhi \cite{Gurzhi-JETP} under the assumption that the bulk momentum relaxation occurs at point like impurities, and the corresponding rate of momentum relaxation is temperature independent. 

The new aspect of modern high-mobility devices is that the correlation radius of the disorder potential $\xi$ may exceed the relaxation length due to momentum-conserving e-e scattering $l_{\text{ee}}$. There are multiple pieces of experimental evidence in support of this. For example, scanning probe microscopy on graphene samples encapsulated by boron nitride shows electron-hole charge puddles with the typical correlation radius in the range of $\xi\sim 100$ nm and local strength of $\delta\mu\sim 5$ meV; see Refs.~\cite{Yacoby-Puddles,Crommie,STM-GhBn}. In the regime where the relaxation length $l_{\text{ee}}$ due to momentum-conserving electron-electron collisions becomes shorter than $\xi$, the disorder-induced momentum relaxation in the bulk may not be described in terms of collisions of individual electrons with impurities, but must be described by using the hydrodynamic approach. For Galilean-invariant electron liquids, this was done in Ref.~\cite{AKS}, and for the electron liquids in graphene, which do not possess charge Galilean invariance, it was done in Refs.~\cite{Lucas,LLA}. In this approach, a description of the system at  spatial scales exceeding the disorder correlation radius is obtained by averaging the hydrodynamic equations in an external potential over disorder realizations.  

\subsection{Disorder-averaged description}

The disorder-averaged Navier-Stokes equation of force balance [Eq.~\eqref{eq:NS-Hall-bar}] sees the following modification: 
\begin{equation}\label{eq:NS-dis}
\eta\partial^2_yu-ku=\vec{x}^{\mathbb{T}}_\text{e}\vec{X}.
\end{equation}  
The key new element here is the appearance of a disorder-induced friction force, which is proportional to the flow velocity (second term in the left-hand side). The corresponding friction coefficient is given by~\cite{LLA}
\begin{equation}\label{eq:k}
k=\frac{\left\langle(s_0\delta n-n_0\delta s)^2\right\rangle}{2\left(\frac{\sigma_0}{e^2}s^2_0-2\frac{\gamma_0}{T}n_0s_0+\frac{\kappa_0}{T}n^2_0\right)}. 
\end{equation} 
Here, subscript zero indicates the values of quantities in the pristine (clean) system, whereas brackets $\langle\ldots\rangle$ denote spatial averages of locally fluctuating particle $\delta n(\bm{r})$ and entropy $\delta s(\bm{r})$ densities. The effective macroscopic densities of particles and entropy, which are transported by the flow, are not equal to the corresponding spatial averages, but are also affected (renormalized) by disorder~\cite{LLA}: $\vec{x}_\text{e}\neq \vec{x}_0$. However, sufficiently close to neutrality and for weak disorder, where $\delta n \ll s_0$, these renormalizations become small and may be neglected. Therefore, below we approximate $\vec{x}_\text{e}\to\vec{x}_0$ in Eq.~\eqref{eq:NS-dis}. Another important change that occurs as compared to the clean limit is the renormalization of matrix $\hat{\Upsilon}$ in Eq.~\eqref{eq:Upsilon}. The most substantial correction appears in the intrinsic conductivity term that is now given by~\cite{LLA}
\begin{equation}\label{eq:chi}
\sigma=\sigma_0+e^2\chi,\quad \chi=\frac{1}{2\eta}\int\frac{d^2q}{(2\pi)^2}\frac{|\delta n(q)|^2}{q^2}.
\end{equation}
The second term above describes the enhancement of the effective conductivity above the intrinsic value. It is caused by disorder-induced deviations from average charge neutrality~\cite{LLA}, and is analogous to the second term in Eq.~\eqref{eq:sigma_clean_inhomogeneous} for clean systems. 

Thus, near charge neutrality, $n_0\ll s_0$, the effect of long-range bulk disorder on the hydrodynamic electron transport can be accounted for by replacing the Navier-Stokes equation with Eq.~(\ref{eq:NS-dis}) and replacing the intrinsic conductivity of the electron liquid with  the macroscopic conductivity $\sigma$ in Eq.~(\ref{eq:chi}).  

For simplicity, we consider the effect of bulk disorder in systems with a uniform macroscopic densities, considered in Sec.~\ref{sec:clean_uniform}. The generalization to systems with inhomogeneous macroscopic density, considered in Sec. \ref{sec:clean_nonuniform}, is straightforward.  Solving Eq.~\eqref{eq:NS-dis} with no-slip boundary conditions, we obtain the hydrodynamic velocity in the sample,
\begin{equation}\label{eq:u-dis}
u(y)=(\vec{x}^{\mathbb{T}}_0\vec{X})\frac{l^2}{\eta}\left[\frac{\cosh(y/l)}{\cosh(d/l)}-1\right],
\end{equation}
where we have introduced a new length scale in the problem,
\begin{equation}\label{eq:Gurzhi-length}
l^{-2}=k/\eta. 
\end{equation}
This length determines the crossover from the bulk to Poiseuille hydrodynamic transport. It is the analog of the Gurzhi length $l_\text{G}$ for the case of point-like impurities, where it is given by the geometric mean of the relaxation lengths due to electron-electron and electron-impurity collisions, $l_\text{G}=\sqrt{l_\text{ee}l_\text{im}}$.   We note that in contrast to the situation with point like impurities, where the rate of momentum-relaxing collisions is independent of e-e scattering, in the present case the rate of momentum relaxation in the bulk, which is characterized by the $k$ in Eq.~\eqref{eq:k}, depends on the rate of e-e scattering. This complicates extraction of the temperature dependence of the viscosity from transport measurements in the regime of Gurzhi crossover (see, also, Ref. \cite{Gurzhi-crossover} for additional discussions).

\subsection{Transport coefficients in the Gurzhi crossover}

With the flow profile in the form of Eq.~\eqref{eq:u-dis}, we evaluate the spatially averaged current from Eq.~\eqref{eq:J} and obtain the resistivity in the form 
\begin{equation}\label{eq:rho_crossover}
\rho_{\text{e}}=\frac{\sigma^{-1}}{1+\frac{e^2}{\sigma}\frac{(n_0l)^2}{\eta}f(d/l)},\quad f(z)=1-\frac{\tanh z}{z}.
\end{equation}
This formula describes the effective resisvity of the system in the entire crossover region between the Poiseuille flow in a clean narrow channel to that of wide channels, where momentum relaxation is dominated by disorder in the bulk. The thermopower can be calculated in the same way, and one finds 
\begin{equation}\label{eq:Q_crossover}
Q_\text{e}=\frac{\gamma_0/T+(n_0s_0l^2/\eta)f(d/l)}{\sigma_0/e^2+\chi+(n^2_0l^2/\eta)f(d/l)}. 
\end{equation}
The expression for the thermal conductivity in the crossover region is
\begin{equation}\label{eq:kappa_crossover}
    \kappa_\text{e} = \kappa+T\frac{(s^2_0l^2/\eta)f(d/l)(\sigma_0/e^2+\chi)}{\sigma_0/e^2+\chi+(n^2_0l^2/\eta)f(d/l)}.
\end{equation}
For narrow channels, $d\ll l$,  Eqs.~\eqref{eq:rho_crossover},  \eqref{eq:Q_crossover}, and \eqref{eq:kappa_crossover} reproduce results for the clean systems, Eqs.~\eqref{eq:rho-clean}, \eqref{eq:Q-clean}, and \eqref{eq:kappa-clean-simple}, respectively.

In the opposite limit of wide channels, $d\gg l$, the system resistivity  in Eq.~\eqref{eq:rho_crossover} becomes independent of the width of $d$, and is given by 
\begin{equation}\label{eq:rho_bulk}
\rho_\text{e}=\frac{\sigma^{-1}_0}{(1+e^2\chi/\sigma_0)+2n^2_0/\langle\delta n^2\rangle}.
\end{equation} 
To arrive at this expression, we additionally approximated $k$ by its value taken close to the charge neutrality $n_0\to0$, where one finds from Eq.~\eqref{eq:k}, that $k=\frac{e^2}{2\sigma_0}\langle\delta n^2\rangle$. 

The thermal conductivity in Eq.~\eqref{eq:kappa_crossover} for wide channels, $d\gg l$,  simplifies to 
\begin{equation}
\label{eq:kappa_bulk}
\kappa_\text{e}=T\frac{2s^2_0}{\langle\delta n^2\rangle}\frac{\sigma_0/e^2+\chi}{(1+e^2\chi/\sigma_0)+2n^2_0/\langle\delta n^2\rangle}.
\end{equation}
Equations \eqref{eq:rho_bulk} and \eqref{eq:kappa_bulk} reproduce the results of Ref.~\cite{LLA}. 

Note that in wide channels,  $d\gg l$, the strong enhancement of the Lorenz ratio still occurs, and the form of the density dependence of $L_\text{e}$ is still given by Eq.~\eqref{eq:L}; however, the width of the Lorentzian peak changes to 
\begin{equation} \label{eq:Gamma_bulk}
\Gamma^2=\frac{1}{2}\left(1+\frac{e^2\chi}{\sigma_0}\right)\langle\delta n^2\rangle.
\end{equation} 

\subsection{Estimates for $\chi$ in graphene systems}
\label{sec:estimates}

The transport coefficients of wide channels [see Eqs.~\eqref{eq:rho_bulk}--\eqref{eq:Gamma_bulk}] depend on the  dimensionless  parameter $\chi$, which determines the magnitude of disorder-induced enhancement of the conductivity above the intrinsic value in Eq.~(\ref{eq:chi}), and is, in general, temperature dependent.  Below, we estimate  $\chi$ for graphene systems near charge neutrality. We assume the density modulations to be weak, $\delta n\ll s_0$, and work in the approximation of linear screening. 

In the hydrodynamic regime, the correlation radius of disorder $\xi$ exceeds the Thomas-Fermi screening radius $a$. The latter is given by $a =1/(2\pi e^2\nu)$, where $\nu=\partial n/\partial\mu\sim T/v^2$ is the thermodynamic density of states. Therefore, the condition $\xi\gg a$ is equivalent to $\xi\gg l_T$, where $l_T=v/T$ is the thermal de Broglie wavelength. Since the interaction constant $e^2/v$ is of the order of unity, this is the applicability condition of the hydrodynamic approximation. Therefore, at length scales of the order of $\xi$, the compressibility of the electron liquid is dominated by the Coulomb interaction. In this case, $\delta n (q)/q =- U(q)/2\pi e^2$, and in  Eq.~(\ref{eq:chi}), we get 
\begin{equation}\label{eq:chi-U}
\chi = \frac{1}{2\eta} \frac{\langle U^2 \rangle}{(2\pi e^2)^2} .
\end{equation}
Note that the temperature dependence of $\chi$ is determined by the temperature dependence of viscosity $\eta$. Near charge neutrality, the latter can be estimated as $\eta\sim (T/v)^2$.  As a consequence, within the window of applicability of the hydrodynamic description, the disorder-induced correction to the  intrinsic conductivity is expected to grow with  decreasing temperature as $\propto 1/T^2$. This may become the dominant source of the temperature dependence of $\sigma$ since the temperature dependence of the intrinsic conductivity is predicted to be logarithmic~\cite{Mishchenko,Kashuba,FSMS}. 

To estimate the magnitude of $\chi$ in Eq.~\eqref{eq:chi-U}, we consider a setup of doping, in which the disorder potential is produced by a layer of dopants with an average density $N_i$ separated from the two-dimensional electron system by a distance $\xi$. In this case, the correlation radius of the disorder potential is set by $\xi$. Assuming the dopants to be spatially uncorrelated, we obtain the spectral power of the external  potential induced in the plane of the electron system in the form
\begin{equation}
\label{eq:U_q}
|U(q)|^2=N_i\left(\frac{2\pi e^2}{q}\right)^2\exp(-2q\xi). 
\end{equation}
Substituting this expression into Eq.~\eqref{eq:chi-U}, we obtain the following estimate for the disorder-induced enhancement of conductivity in Eq.~\eqref{eq:chi}:
\begin{equation}\label{eq:sigma-visc-correction}
\delta \sigma=e^2\chi\simeq\frac{e^2N_i}{4\pi\eta}\ln\frac{L}{\xi}. 
\end{equation}
Here the infrared logarithmic divergence of the $q$ integral was regularized by the system size $L$. This divergence arises from the assumption of Poissonian distribution of dopants, which leads to the divergence of the spectral power of disorder in Eq.~\eqref{eq:U_q} at $q\to 0$.  This estimate agrees with the logarithmic renormalization term in Eq.~(4) of Ref.~\cite{Hruska} (see, also, discussions that lead to Eq.~(5.1) in Ref.~\cite{Pal} and more recent work~\cite{Guo-2}).

For sources of long-range disorder with spectral density that does not have
strong divergence at $q\to 0$ (e.g., encapsulation-induced disorder), Eq. \eqref{eq:chi-U} yields the estimate $\chi \sim (v/e^2)^2\frac{\langle U^2 \rangle}{T^2}$, where we took $\eta\sim (T/v)^2$. 


\section{Summary}

In this work, we studied hydrodynamic thermoelectric transport in Hall-bar devices with non-Galilean-invariant electron liquids. We  focused on the effects of inhomogeneity of electron density induced by the electrostatic gates and/or long-range disorder potential. We showed that spatial density variations in the sample lead to strong mixing between thermoelectric transport relative the electron liquid and convective transport of charge and energy by the hydrodynamic flow. In particular, in nominally charge-neutral graphene devices, charge transport is generally strongly affected by the hydrodynamic flow. This leads to strong enhancement of the effective conductivity over the intrinsic value. 

We evaluated the electrical resistivity, thermal conductivity, as well as the Seebeck and Peltier coefficients of Hall-bar systems. In the clean limit for devices with uniform electron density, these are given by  Eqs. \eqref{eq:rho-clean}--\eqref{eq:kappa-clean}. In clean devices with nonuniform gate-induced density, they are given by Eq.~\eqref{eq:transport_clean_inhomogeneous}. 

In Sec.~\ref{sec:disorder}, we studied the effect of the long-range disorder potential on the electron transport and obtained the thermoelectric coefficients of the systems in the full crossover range between the clean and disorder-dominated regimes; see Eqs.~\eqref{eq:rho_crossover}--\eqref{eq:kappa_crossover}. 
The crossover occurs when the Hall-bar width becomes comparable to $l$ in
Eq. \eqref{eq:Gurzhi-length}. In the case of a long-range disorder, this length is defined not only by the disorder strength, but also the viscosity and  the entire matrix of intrinsic kinetic coefficients of the electron liquid, via Eq. \eqref{eq:k}. It is the analog of the Gurzhi length $l_{\text{G}} =\sqrt{l_{\text{ee}}l_{\text{im}}}$.
 

\subsection*{Acknowledgments}

We thank S. Ilani, G. Falkovich, M. Foster, P. Kim, L. Levitov, A. Talanov, and J. Waissman for insightful discussions. We also thank A. Principi for the communications regarding Ref. \cite{Principi-2DM}. 

This work was financially supported by the U.S. Department of Energy, Office of Science, Basic Energy Sciences Program for Materials and Chemistry Research in Quantum Information Science under Award No. DE-SC0020313 (A. L.), National Science Foundation Grant No. DMR-1653661 (S. L.),  and MRSEC Grant No. DMR-1719797 (A.V.A.). This project was initiated during the workshop ``From Chaos to Hydrodynamics in Quantum Matter" at the Aspen Center for Physics, which is supported by the National Science Foundation Grant No. PHY-1607611.


\begin{thebibliography}{99}

\bibitem{Spivak-RMP}
B. Spivak, S. V. Kravchenko, S. A. Kivelson, and X. P. A. Gao, 
\textit{Colloquium: Transport in strongly correlated two dimensional electron fluids}, 
Rev. Mod. Phys. \textbf{82}, 1743 (2010). 

\bibitem{NGMS}
B. N. Narozhny, I. V. Gornyi, A. D. Mirlin, J. Schmalian,
\textit{Hydrodynamic Approach to Electronic Transport in Graphene},
Ann. Phys. \textbf{529}, 1700043 (2017).

\bibitem{Lucas-Fong}
A. Lucas, K.C. Fong,
\textit{Hydrodynamics of electrons in graphene},
J. Phys.: Condens. Matter \textbf{30}, 053001 (2018).

\bibitem{Polini-Geim}
Marco Polini, Andre K. Geim, 
\textit{Viscous electron fluids},
Physics Today \textbf{73}, 28 (2020).

\bibitem{ALJS}
Alex Levchenko and J\"org Schmalian,
\textit{Transport properties of strongly coupled electron–phonon liquids},
Annals of Physics \textbf{419}, 168218 (2020).

\bibitem{Gurzhi-JETP}
R. N. Gurzhi, 
\textit{Minimum of Resistance in Impurity-free Conductors}, 
Sov. JETP \textbf{17}, 521 (1963).

\bibitem{Gurzhi-UFN}
R. N. Gurzhi, \textit{Hydrodynamic effects in solids at low temperature}, 
Sov. Phys. Usp. \textbf{11}, 255 (1968).

\bibitem{LWM1}
L. W. Molenkamp, M. J. M. de Jong, 
\textit{Electron-electron-scattering-induced size effects in a two-dimensional wire}, 
Phys. Rev. B \textbf{49}, 5038 (1994).

\bibitem{LWM2}
M. J. M. de Jong, L. W. Molenkamp, \textit{Hydrodynamic electron flow in high-mobility wires}, 
Phys. Rev. B \textbf{51}, 13389 (1995). 

\bibitem{Abrikos-Khalat}
A. A. Abrikosov and I. M. Khalatnikov, \textit{The theory of a Fermi liquid}, Rep. Prog. Phys. \textbf{22},
329 (1959).

\bibitem{Abrikosov-Book}
A. A. Abrikosov, 
\textit{Fundamentals of the Theory of Metals}, Dover Publications; Reprint edition (October 18, 2017). 

\bibitem{AKS}
A. V. Andreev, Steven A. Kivelson, B. Spivak, 
\textit{Hydrodynamic description of transport in strongly correlated electron systems}, 
Phys. Rev. Lett. \textbf{106}, 256804 (2011).

\bibitem{Fritz}
Markus M\"uller, J\"{o}rg Schmalian, and Lars Fritz,
\textit{Graphene: A Nearly Perfect Fluid},
Phys. Rev. Lett. \textbf{103}, 025301 (2009).

\bibitem{Torre}
Iacopo Torre, Andrea Tomadin, Andre K. Geim, and Marco Polini, 
\textit{Nonlocal transport and the hydrodynamic shear viscosity in graphene}, Phys. Rev. B \textbf{92}, 165433 (2015). 

\bibitem{Levitov-Falkovich}
Leonid Levitov and Gregory Falkovich, 
\textit{Electron viscosity, current vortices and negative nonlocal resistance in graphene},
Nature Physics \textbf{12}, 672 (2016).

\bibitem{Falkovich-Levitov}
Gregory Falkovich and Leonid Levitov, 
\textit{Linking Spatial Distributions of Potential and Current in Viscous Electronics}, 
Phys. Rev. Lett. \textbf{119}, 066601 (2017). 

\bibitem{Bandurin-1}
D. A. Bandurin, I. Torre, R. K. Kumar, M. B. Shalom, A. Tomadin, A. Principi, G. H. Auton, E. Khestanova, K. S. NovoseIov, I. V. Grigorieva, L. A. Ponomarenko, A. K. Geim, M. Polini,
\textit{Negative local resistance due to viscous electron backflow in graphene},
Science \textbf{351}, 1055 (2016).

\bibitem{Bandurin-2}
D. A. Bandurin, A. V. Shytov, L. Levitov, R. K. Kumar, A. I. Berdyugin, M. Ben-Shalom, I. V. Grigorieva, A. K. Geim, G. Falkovich,
\textit{Fluidity onset in graphene},
Nat. Commun. \textbf{9}, 4533 (2018).

\bibitem{Zeldov}
Amit Aharon-Steinberg, Tobias V\"olkl, Arkady Kaplan, Arnab K. Pariari, Indranil Roy, Tobias Holder, Yotam Wolf, Alexander Y. Meltzer, Yuri Myasoedov, Martin E. Huber, Binghai Yan, Gregory Falkovich, Leonid S. Levitov, Markus H\"ucker, Eli Zeldov, \textit{Direct observation of vortices in an electron fluid}, arXiv:2202.02798 [cond-mat.mes-hall]. 

\bibitem{Ensslin}
B. A. Braem, F. M. D. Pellegrino, A. Principi, M. R\"{o}\"o{}sli, C. Gold, S. Hennel, J. V. Koski, M. Berl, W. Dietsche, W. Wegscheider, M. Polini, T. Ihn, and K. Ensslin, \textit{Scanning gate microscopy in a viscous electron fluid}, Phys. Rev. B \textbf{98}, 241304(R) (2018).

\bibitem{Sulpizio} 
J.A. Sulpizio, L. Ella, A. Rozen, J. Birkbeck, D.J. Perello, D. Dutta, M. Ben-Shalom, T. Taniguchi, K. Watanabe, T. Holder,
R. Queiroz, A. Stern, T. Scaffidi, A.K. Geim, S. Ilani, \textit{Visualizing Poiseuille flow of hydrodynamic electrons}, 
Nature \textbf{576}, 75 (2019).

\bibitem{Ku}
M. J. H. Ku, T. X. Zhou, Q. Li, Y. J. Shin, J. K. Shi, C. Burch, H. Zhang, F. Casola, T. Taniguchi, K. Watanabe, P. Kim, A. Yacoby and R. L. Walsworth, \textit{Imaging Viscous Flow of the Dirac Fluid in Graphene Using a Quantum Spin Magnetometer}, Nature \textbf{583}, 537 (2020).

\bibitem{Jenkins}
A. Jenkins, S. Baumann, H. Zhou, S. A. Meynell, D. Yang, K. Watanabe, T. Taniguchi, A. Lucas, A. F. Young and A. C. Bleszynski Jayich, \textit{Imaging the breakdown of ohmic transport in graphene}, arXiv:2002.05065 [cond-mat.mes-hall].

\bibitem{Ilani}
Chandan Kumar, John Birkbeck, Joseph A. Sulpizio, David J. Perello, Takashi Taniguchi, Kenji Watanabe, Oren Reuven, Thomas Scaffidi, Ady Stern, Andre K. Geim, Shahal Ilani, \textit{Imaging Hydrodynamic Electrons Flowing Without Landauer-Sharvin Resistance}, 
arXiv:2111.06412 [cond-mat.mes-hall].

\bibitem{Guo-1}
Haoyu Guo, Ekin Ilseven, Gregory Falkovich, Leonid Levitov, 
\textit{Higher-Than-Ballistic Conduction of Viscous Electron Flows}, 
PNAS \textbf{114}, 3068 (2017).

\bibitem{Kumar}
R. K. Kumar, D. A. Bandurin, F. M. D. Pellegrino, Y. Cao, A. Principi, H. Guo, G. H. Auton, M. Ben-Shalom, L. A. Ponomarenko, G. Falkovich, K. Watanabe, T. Taniguchi, I. V. Grigorieva, L. S. Levitov, M. Polini, A. K. Geim,
\textit{Superballistic flow of viscous electron fluid through graphene constrictions},
Nat. Phys. \textbf{13}, 1182 (2017).

\bibitem{Brar}
Zachary Krebs, Wyatt Behn, Songci Li, Keenan Smith, Kenji Watanabe, Takashi Taniguchi, Alex Levchenko, Victor Brar,
\textit{Imaging the breaking of electrostatic dams in graphene for ballistic and viscous fluids},
arXiv:2106.07212 [cond-mat.mes-hall].

\bibitem{Hruska}
M. Hruska and B. Spivak, Phys. Rev. B \textbf{65}, 033315 (2002).

\bibitem{Pal}
H. K. Pal, V. I. Yudson, D. L. Maslov, 
\textit{Resistivity of non-Galilean-invariant Fermi- and non-Fermi liquids}, 
Lith. J. Phys. \textbf{52}, 142 (2012).

\bibitem{Guo-2}
Haoyu Guo, Ekin Ilseven, Gregory Falkovich, Leonid Levitov, 
\textit{Stokes Paradox, Back Reflections and Interaction-Enhanced Conduction}, arXiv:1612.09239 [cond-mat.mes-hall]. 

\bibitem{LLA}
Songci Li, Alex Levchenko, and A. V. Andreev,
\textit{Hydrodynamic electron transport near charge neutrality},
Phys. Rev. B \textbf{102}, 075305 (2020).

\bibitem{Aleiner}
M. S. Foster and I. L. Aleiner,
\textit{Slow imbalance relaxation and thermoelectric transport in graphene},
Phys. Rev. B \textbf{79}, 085415 (2009).

\bibitem{NGTSM}
B. N. Narozhny, I. V. Gornyi, M. Titov, M. Sch\"utt, A. D. Mirlin,
\textit{Hydrodynamics in graphene: Linear-response transport},
Phys. Rev. B \textbf{91}, 035414 (2015).

\bibitem{Lucas}
A. Lucas, J. Crossno, K. C. Fong, P. Kim, and S. Sachdev,
\textit{Transport in inhomogeneous quantum critical fluids and in the Dirac fluid in graphene},
Phys. Rev. B \textbf{93}, 075426 (2016).

\bibitem{Xie-Foster}
H.-Y. Xie and M. S. Foster,
\textit{Transport coefficients of graphene: Interplay of impurity scattering, Coulomb interaction, and optical phonons},
Phys. Rev. B \textbf{93}, 195103 (2016).

\bibitem{Principi-2DM}
M. Zarenia, A. Principi, G. Vignale,
\textit{Disorder-enabled hydrodynamics of charge and heat transport in monolayer graphene},
2D Mater. \textbf{6}, 035024 (2019).

\bibitem{Narozhny-Gornyi}
B. N. Narozhny, I. V. Gornyi, \textit{Hydrodynamic approach to electronic transport in graphene: energy relaxation}, 
Frontiers in Physics \textbf{9}, 640649 (2021).

\bibitem{Crossno}
J. Crossno, J. K. Shi, K. Wang, X. Liu, A. Harzheim, A. Lucas, S. Sachdev, P. Kim, T. Taniguchi, K. Watanabe, T. A. Ohki, K. C. Fong,
\textit{Observation of the Dirac fluid and the breakdown of the Wiedemann-Franz law in graphene},
Science \textbf{351}, 1058 (2016).

\bibitem{Ghahari}
F. Ghahari, H.-Y. Xie, T. Taniguchi, K. Watanabe, M. S. Foster, and P. Kim,
\textit{Enhanced Thermoelectric Power in Graphene: Violation of the Mott Relation by Inelastic Scattering},
Phys. Rev. Lett. \textbf{116}, 136802 (2016).

\bibitem{Smet}
Johannes Geurs, Youngwook Kim, Kenji Watanabe, Takashi Taniguchi, Pilkyung Moon, Jurgen H. Smet, 
\textit{Rectification by hydrodynamic flow in an encapsulated graphene Tesla valve}, arXiv:2008.04862 [cond-mat.mes-hall].  

\bibitem{Hamilton}
Aydin Cem Keser, Daisy Q. Wang, Oleh Klochan, Derek Y. H. Ho, Olga A. Tkachenko, Vitaly A. Tkachenko, Dimitrie Culcer, Shaffique Adam, Ian Farrer, David A. Ritchie, Oleg P. Sushkov, and Alexander R. Hamilton, \textit{Geometric Control of Universal Hydrodynamic Flow in a Two-Dimensional Electron Fluid}, 
Phys. Rev. X \textbf{11}, 031030 (2021).

\bibitem{Shavit}
Michal Shavit, Andrey Shytov, and Gregory Falkovich,
\textit{Freely Flowing Currents and Electric Field Expulsion in Viscous Electronics},
Phys. Rev. Lett. \textbf{123}, 026801 (2019).

\bibitem{LLA-Corbino}
Songci Li, Alex Levchenko, A. V. Andreev, 
\textit{Hydrodynamic thermoelectric transport in Corbino geometry}, 
Phys. Rev. B \textbf{105}, 125302 (2022). 

\bibitem{LL-V6}
L. D. Landau and E. M. Lifshitz,
\textit{Fluid Mechanics}: Volume 6 Course of Theoretical Physics Series (Butterworth-Heinemann, 2nd edition, 1987).

\bibitem{LL-V5}
L. D. Landau and E. M. Lifshitz,
\textit{Statistical Physics}:
Volume 5 of Course of Theoretical Physics Series
(Butterworth-Heinemann, Oxford, 2013), 3$^{\text{rd}}$ edition.

\bibitem{Mishchenko}
E. G. Mishchenko, 
\textit{Effect of Electron-Electron Interactions on the Conductivity of Clean Graphene}, 
Phys. Rev. Lett. \textbf{98}, 216801 (2007). 

\bibitem{Kashuba}
Alexander B. Kashuba, 
\textit{Conductivity of defectless graphene}, 
Phys. Rev. B \textbf{78}, 085415 (2008).

\bibitem{FSMS}
Lars Fritz, J\"org Schmalian, Markus M\"uller, and Subir Sachdev,
\textit{Quantum critical transport in clean graphene},
Phys. Rev. B \textbf{78}, 085416 (2008).

\bibitem{Zeldov-Nature}
Dorri Halbertal, Jo Cuppens, Moshe Ben Shalom, Lior Embon, Nitzan Shadmi, Yonathan Anahory, HR Naren, Jayanta Sarkar, Aviram Uri, Yuval Ronen, Yury Myasoedov, Leonid Levitov, Ernesto Joselevich, Andre Konstantin Geim, Eli Zeldov, \textit{Nanoscale thermal imaging of dissipation in quantum systems}, Nature \textbf{539}, 407 (2016).

\bibitem{Zeldov-Science}
Dorri Halbertal, Moshe Ben Shalom, Aviram Uri, Kousik Bagani, Alexander Y. Meltzer, Ido Marcus, Yuri Myasoedov, John Birkbeck, Leonid S. Levitov, Andre K. Geim, Eli Zeldov, \textit{Imaging resonant dissipation from individual atomic defects in graphene}, Science \textbf{358}, 1303 (2017).

\bibitem{Waissman}
Jonah Waissman, Laurel E. Anderson, Artem V. Talanov, Zhongying Yan, Young J. Shin, Danial H. Najafabadi, Mehdi Rezaee, Xiaowen Feng, Daniel G. Nocera, Takashi Taniguchi, Kenji Watanabe, Brian Skinner, Konstantin A. Matveev, Philip Kim, \textit{Electronic Thermal Transport Measurement in Low-Dimensional Materials with Graphene Nonlocal Noise Thermometry}, 
Nature Nanotechnology \textbf{17}, 166 (2022).

\bibitem{Yacoby-Puddles} 
J. Martin, N. Akerman, G. Ulbricht, T. Lohmann, J. H. Smet, K. von Klitzing, and A. Yacoby, 
\textit{Observation of electron-hole puddles in graphene using a scanning single-electron transistor}, 
Nat. Phys. \textbf{4}, 144 (2008).

\bibitem{Crommie}
Y. Zhang, V. W. Brar, C. Girit, A. Zettl, and M. F. Crommie,
\textit{Origin of spatial charge inhomogeneity in graphene},
Nat. Phys. \textbf{5}, 722 (2009).

\bibitem{STM-GhBn}
Jiamin Xue, Javier Sanchez-Yamagishi, Danny Bulmash, Philippe Jacquod, Aparna Deshpande, K. Watanabe, T. Taniguchi, Pablo Jarillo-Herrero and Brian J. LeRoy \textit{Scanning tunnelling microscopy and spectroscopy of ultra-flat graphene on hexagonal boron
nitride}, Nat. Mater. \textbf{10}, 282 (2011).

\bibitem{Gurzhi-crossover}
Songci Li, Maxim Khodas, Alex Levchenko, 
\textit{Conformal maps of viscous electron flow in the Gurzhi crossover}, 
Phys. Rev. B \textbf{104}, 155305 (2021).

\end{thebibliography}
\end{document}